\documentclass[12pt]{article}
\usepackage{graphicx}

\newcommand\pubnumber{DPF2013-142}
\newcommand\pubdate{\today}

\def\Title#1{\begin{center} {\Large #1 } \end{center}}
\def\Author#1{\begin{center}{ \sc #1} \end{center}}
\def\Address#1{\begin{center}{ \it #1} \end{center}}

\newcommand\pubblock{\rightline{\begin{tabular}{l} \pubnumber\\
         \pubdate  \end{tabular}}}
\newenvironment{Abstract}{\begin{quotation}  }{\end{quotation}}
\newenvironment{Presented}{\begin{quotation} \begin{center} 
             PRESENTED AT\end{center}\bigskip 
      \begin{center}\begin{large}}{\end{large}\end{center} \end{quotation}}

\def\beq{\begin{equation}}
\def\eeq#1{\label{#1}\end{equation}}
\def\eeqn{\end{equation}}

\def\beqa{\begin{eqnarray}}
\def\eeqa#1{\label{#1}\end{eqnarray}}
\def\eeqan{\end{eqnarray}}

\let\bar=\overbar

\def\Dslash{\not{\hbox{\kern-4pt $D$}}}
\def\dslash{\not{\hbox{\kern-2pt $\del$}}}

\def\msb{{\bar{\ssstyle M \kern -1pt S}}}

\begin{document}
\begin{titlepage}
\pubblock

\vfill
\Title{Particle identification with the iTOP detector at Belle-II}
\vfill
\Author{Matthew Barrett\\
        University of Hawai`i at M\={a}noa\\
        On behalf of the Belle-II iTOP group}
\Address{2505 Correa Road, Honolulu, HI 96822\\United States of America}
\vfill
\begin{Abstract}
The Belle-II experiment and superKEKB accelerator will form a next generation B-factory at KEK, capable of running at an instantaneous luminosity 40 times higher than the Belle detector and KEKB. This will allow for the elucidation of many facets of the Standard Model by performing precision measurements of its parameters, and provide sensitivity to many rare decays that are currently inaccessible. This will require major upgrades to both the accelerator and detector subsystems. The imaging Time-of-propagation (iTOP) detector will be a new subdetector of Belle-II that will perform an integral role in Particle identification (PID). It will comprise 16 modules between the tracking detectors and calorimeter; each module consisting of a quartz radiator, approximately 2.5m in length, instrumented with an array of 32 micro-channel plate photodetectors (MCP-PMTs). The passage of charged particles through the quartz will produce a cone of Cherenkov photons that will propagate along the length of the quartz, and be detected by the MCP-PMTs. The excellent spatial, and timing resolution (of 50 picoseconds) of the iTOP system will provide superior particle identification capabilities, particularly allowing for enhanced discrimination between pions and kaons that will be essential for many of the key measurements to performed. The status of the construction of the iTOP subdetector, and performance studies of prototypes at beam tests will be presented, together with prospects for physics measurements that will utilise the PID capabilities of the iTOP system.
\end{Abstract}
\vfill
\begin{Presented}
DPF 2013\\
The Meeting of the American Physical Society\\
Division of Particles and Fields\\
Santa Cruz, California, August 13--17, 2013\\
\end{Presented}
\vfill
\end{titlepage}
\def\thefootnote{\fnsymbol{footnote}}
\setcounter{footnote}{0}

\section{From Belle to Belle-II PID}

The Belle detector~\cite{belle} collected data for over ten years from 1999 onwards, recording electron-positron collisions produced at the KEKB accelerator, and accumulating over one inverse attobarn of data.  Belle-II will be an upgrade of the Belle detector, that will run in conjunction with superKEKB, an upgrade of the KEKB accelerator.  The luminosity, both instantaneous and integrated, for Belle-II/superKEKB will be a factor of 40-50 times higher than for their predecessors. 

To take data in this higher luminosity environment many of the subdectors from Belle require major upgrades or replacement.  Particle Identification (PID) at Belle was achieved using a time-of-flight counter, and an aerogel Cherenkov counter.  The latter used several layers of aerogel with differing refractive indices to detect differences in the Cherenkov light emitted by particles of different masses.  

In Belle-II PID will be performed by two dedicated subsystems:  For the barrel region of the detector the imaging Time-Of-Propagation (iTOP) subdetector will be used, with an Aerogel Ring Imaging Cherenkov (ARICH) subdetector used in the endcap regions.  Together with tracking and other information, such as energy loss ($dE/dx$), from other subdetectors these dedicated PID systems will allow for the discrimination between different types of particle, and identification of such.  It is of paramount importance to have discriminating power between charged pions and kaons for many of the physics analyses which will be performed at Belle-II, which will be provided by the iTOP and ARICH.  They will also allow for the identification of other charged particles, such as electrons(positrons), protons, and muons.

This document describes the principle of operation, and the design considerations of the iTOP subdetector, together with some early results of testing an iTOP prototype.

\section{The iTOP Detector}

\begin{figure}[htb]
\centering
\includegraphics[width=\columnwidth]{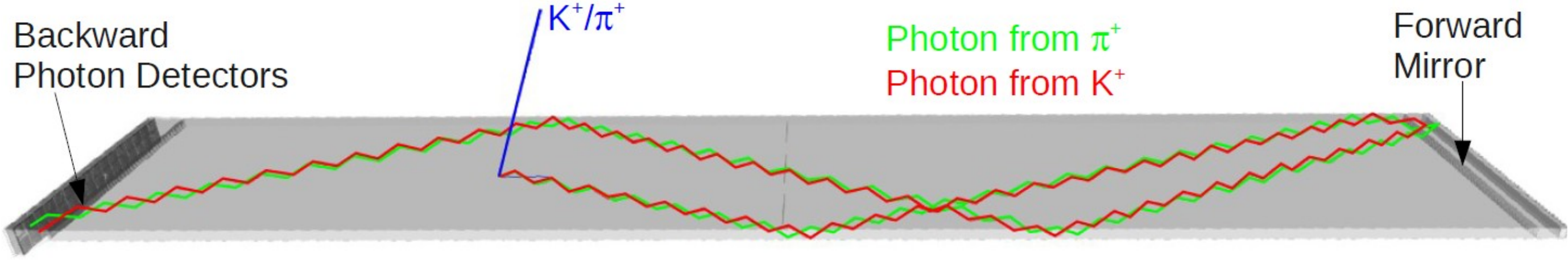}
\caption{Diagram showing two photons emitted during the passage of a charged particle through the quartz of an iTOP module.  One of the photons is emitted at an angle corresponding to the charged particle being a kaon, and the other photon is emitted at an angle corresponding to the charged particle being a pion.  The arrival time and arrival position will be different depending on whether the charged particle was a pion or a kaon.}
\label{fig:Operation}
\end{figure}

In Belle-II there will be sixteen iTOP modules forming a barrel.  A diagram of a single module is shown in Figure~\ref{fig:Operation}.  Each of the modules has two quartz bars, with a total length of approximately 2.5 metres.  At the forward end of the module is a mirror, which will reflect any photons that hit it.  At the backward end of the module is an expansion prism coupled an array of pixelated photodetectors and the associated readout electronics.

When a charged particle passes through a material at greater than the local phase velocity of light it emits Cherenkov radiation~\cite{cherenkov}.  The cosine of the angle at which the Cherenkov photons are emitted is inversely proportional to the speed of the particle, and the refractive index of the material.  Thus photons are emitted at different angles for a given particle track, depending on the mass of the particle that created the track.  

In the iTOP module Chernkov photons are created in the quartz bars, and then propagate through the bars, reflecting from the surfaces of the quartz because of total internal refection.
This is shown in Figure~\ref{fig:Operation} for an iTOP module for the cases when the charged particle is either a pion or a kaon.

For a given track the distribution of photons that are detected by the photodetectors will be different depending on the particle that produced the track.  A photon hitting a particular location on the photodetectors that originated from a kaon would, on average, arrive at a later time than one that originated from a pion.  Each track typically produces twenty to thirty detected Cherenkov photons.  The location and recorded time of each of the detected Cherenkov photons can be combined together to form a likelihood from given particle hypotheses, with the most likely particle identification assigned to the track.  To ensure optimal performance of the iTOP excellent time resolution is required for the detected photons.  

\section{Design Considerations}

The full Belle-II detector will require 32 quartz bars, each $20$mm$\times450$mm$\times1250$mm, with two bars per module.  The quartz needs to be of high quantity to ensure that any photon losses during propagation through the bar are small, and that the Cherenkov photons maintain their reflection angle though many reflections from the surfaces of the quartz.  Table~\ref{tab:quartz} shows some of the requirements placed on the properties of the quartz bars to ensure the performance of the modules.  Similar requirements are placed on the quartz of the short bars with the mirror, and the expansion prism.

\begin{table}[htb]
\begin{center}
\begin{tabular}{|l|l|}  
\hline
Quartz property &  Requirement \\ \hline
Flatness & $<6.3$$\mu$m \\
Perpendicularity & $<20$ arcsec \\
Parallelism & $< 4$ arcsec \\
Roughness & $<0.5$nm (RMS) \\
Bulk transmittance & $>98$\%/m \\
Surface reflectance & $>99.9$\%/reflection \\ \hline
\end{tabular}
\caption{Requirements placed on the quartz bars of the iTOP modules.}
\label{tab:quartz}
\end{center}
\end{table}

To detect the Cherenkov photons each module contains 32 Micro-Channel-Plate Photomultiplier Tubes (MCP-PMT), in two rows of 16.  Each MCP-PMT has an active area of approximately 22.5mm$\times$22.5mm, composed of an NaKSbCs photocathode.  For readout each PMT anode is divided into 4$\times$4 channels, giving a total of 512 channels per iTOP module.  To ensure effective performance each PMT is required to have a peak quantum efficiency of $>$24\%, together with a collection efficiency of 55\%.  These PMTs has a gain of $\sim$1$\times10^{6}$ at the nominal operating high voltage, and an intrinsic transit time spread of $\sim$40ps.

The readout system, based on switched capacitor arrays, for the signals recorded in the PMTs includes a number of waveform sampling ASICs (Application-Specific Integrated Circuit), with two ASICs used to read out each PMT.  The charge deposited on a PMT anode is converted to a waveform that is used to determine the photon detection time, with a time resolution of 50ps.  In order to achieve this resolution each of the ASICs needs to undergo extensive calibration.

\section{Beam test at SPring-8}

In June 2013 a beam test using a prototype iTOP module was undertaken at the LEPS beamline at the SPring-8 laboratory in Japan.   The LEPS experiment uses a $\gamma$ beam of up to 2.4GeV produced by inverse Compton scattering. For the beam test this was collided with a lead target to produce electrons and positrons.  The LEPS magnetic field, together with a triggering system was used to select a beam of postitrons with a mean energy of 2.1GeV.  This positron beam was then used in a number of different setups, each simulating possible incident positions, and associated angles, on the module, such as would be expected in Belle-II.  Tracking and momentum information was provided by the LEPS subdetectors, and small scintillating fibre trackers placed before and after the module, athwart the positron beam.

\begin{figure}[htb]
\centering
\includegraphics[width=\columnwidth]{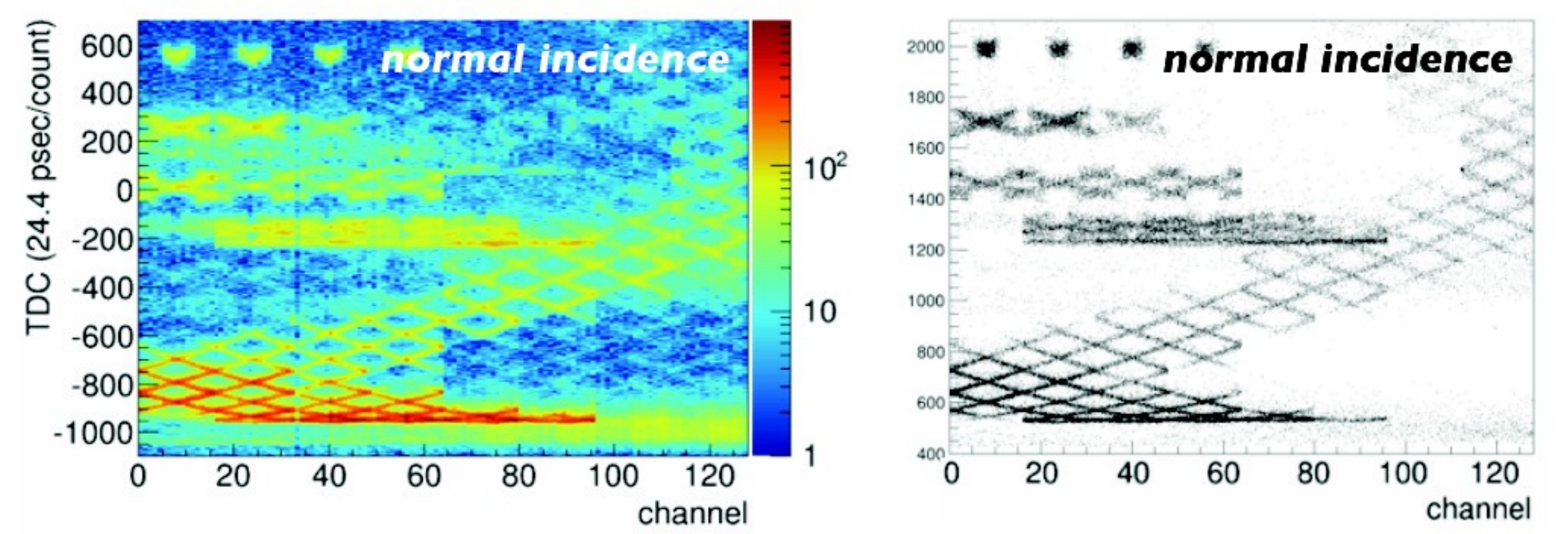}
\caption{The ``ring image'' for the iTOP prototype module using constant fraction discriminator electronics for the experimental set-up where the positron beam hit the prototype module at normal incidence.  The left-hand image is obtained from data, and the right-hand image shows the prediction from simulations.  With these electronics groups of four channels were read out as a single channel, hence a total of 128 channels.}
\label{fig:CFDring}
\end{figure}

Data were taken with the beam hitting the quartz at normal incidence, and at forward angles; the majority of the detected photons will have reflected from the iTOP mirror in the latter set-up.
   
Two sets of readout electronics were used.  Firstly IRS3B waveform sampling ASIC based electronics; the results from these electronics require extensive calibration to achieve the desired time resolution.  Secondly constant fraction discriminator electronics were used; Figure~\ref{fig:CFDring} shows a ``ring image'' produced with these electronics.  The ring image shows the projection of the original ring of Cherenkov photons that were detected, both temporally and spatially (described via the sequentially numbered channels readout from the MCP-PMTs).

\begin{figure}[htb!]
\centering
\includegraphics[width=\columnwidth]{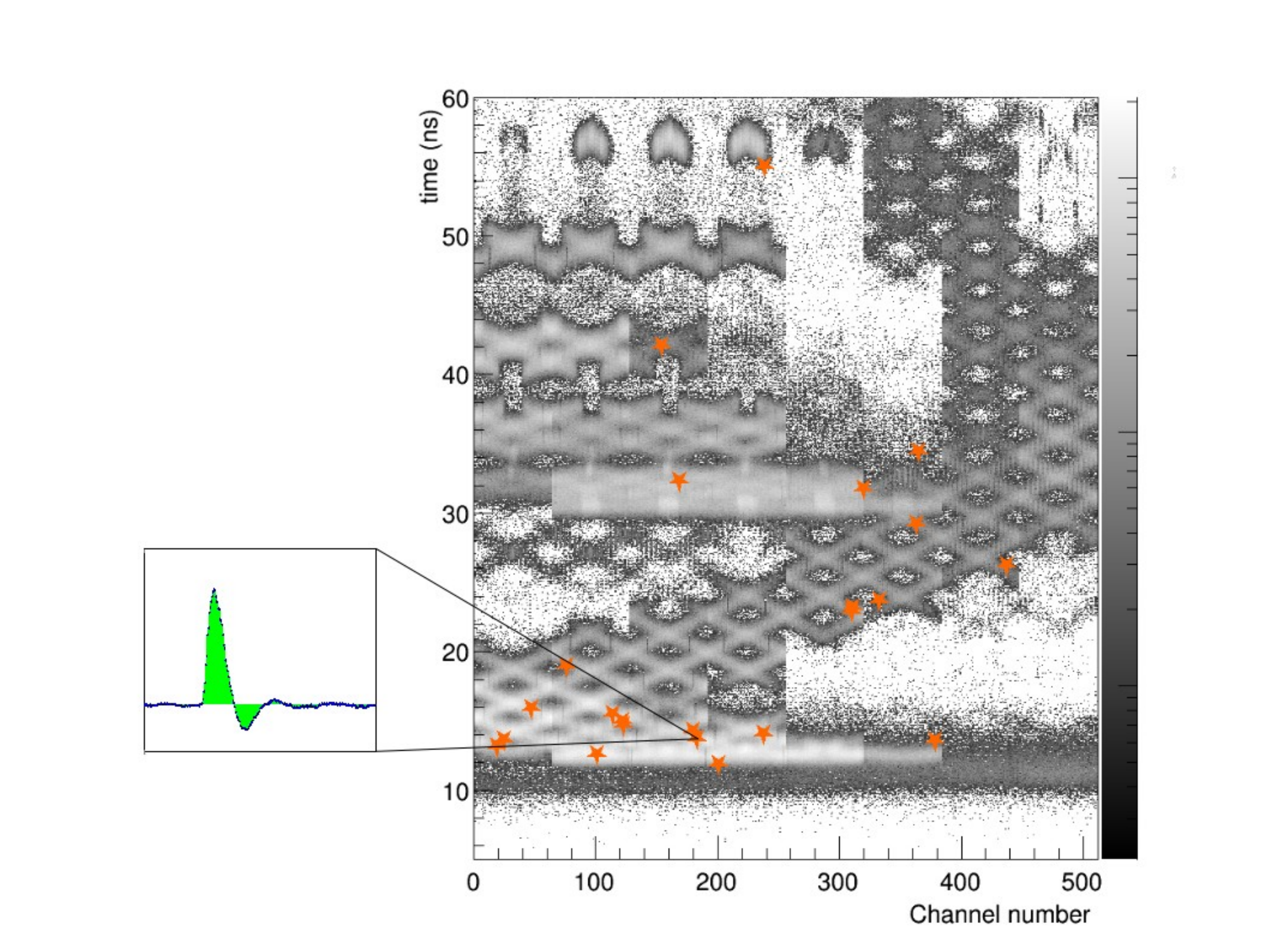}
\caption{A single event recorded with the IRS3B electronics.  Each Cherenkov photon is detected via the charge deposited on a PMT anode as a waveform recorded with the IRS3B ASIC.  An example waveform is shown in the inset image on the left-hand side.  Each waveform is converted to a time; the time and  location for the photons detected in a single event are shown here (stars), superimposed upon the predicted distribution from many simulated events (greyscale bands).}
\label{fig:MarlowExample}
\end{figure}

The ring image is integrated over many events to produce the bands that can be seen.  A single event, containing 20 to 30 detected photons will not show the distinct features visible in the integrated ring image.  An example of a single event recorded at the beam test is shown in Figure~\ref{fig:MarlowExample}.  During the operation of Belle-II the distributions of photons from a single event will be compared with the predicted distributions for different particle hypotheses (pion, kaon, proton, ...) to determine particle likelihoods, which will be combined with additional information from other subdetectors to perform Belle-II's particle identification.


\section{Summary}
The iTOP detector will perform particle identification in the barrel region of Belle-II.  It is of great importance that it be able to distinguish between pions and kaons.  To perform this task excellent time resolution of the Cherenkov photons detected from the passage of charged particles is required.
A prototype of the iTOP module was recently tested at a beam test, with initial analysis showing good agreement between the data taken, and predictions from simulation. 


\end{document}